\documentclass{article}
\usepackage{waspaa23,amsmath,graphicx,url,times}
\usepackage{color}
\usepackage{multirow}
\usepackage{booktabs}
\usepackage{hyperref}
\usepackage{cite}
\usepackage{floatrow}
\usepackage{amssymb}

\usepackage[T1]{fontenc}
\usepackage[utf8]{inputenc}
\usepackage{multicol}
\usepackage{xcolor}
\usepackage{lineno}

\newfloatcommand{capbtabbox}{table}[][\FBwidth]

\usepackage[square, numbers,sort&compress]{natbib}
\renewcommand{\refname}{REFERENCES}
\makeatletter
\renewcommand{\bibsection}{
  \section{\refname
            \@mkboth{\MakeUppercase{\refname}}{\MakeUppercase{\refname}}
  }\vspace{-0.4cm}
}

\title{Music De-limiter Networks via Sample-wise Gain Inversion}

\name{Chang-Bin Jeon$^{1}$, Kyogu Lee$^{1,2,3}$}
\address{$^1$Department of Intelligence and Information, Seoul National University\\
            $^2$Interdisciplinary Program in Artificial Intelligence, Seoul National University\\
            $^3$AI Institute, Seoul National University, Seoul, South Korea
            }

\begin{document}

\setlength{\aboverulesep}{0pt}
\setlength{\belowrulesep}{0pt}
\setlength{\textfloatsep}{5pt}
\setlength{\abovecaptionskip}{0pt}
\setlength{\belowcaptionskip}{-10pt}

\ninept
\maketitle

\begin{sloppy}

\begin{abstract}
The loudness war, an ongoing phenomenon in the music industry characterized by the increasing final loudness of music while reducing its dynamic range,
has been a controversial topic for decades. Music mastering engineers have used \textit{limiters} to heavily compress and make music louder, which can induce ear fatigue and hearing loss in listeners. In this paper, we introduce music \textit{de-limiter} networks that estimate uncompressed music from heavily compressed signals. Inspired by the principle of a limiter, which performs sample-wise gain reduction of a given signal, we propose the framework of \textit{sample-wise gain inversion} (SGI). We also present the \textit{musdb-XL-train} dataset, consisting of 300k segments created by applying a commercial limiter plug-in for training real-world friendly de-limiter networks. Our proposed de-limiter network achieves excellent performance with a scale-invariant source-to-distortion ratio (SI-SDR) of
24.0 dB
in reconstructing \textit{musdb-HQ} from \textit{musdb-XL} data, a limiter-applied version of \textit{musdb-HQ}. The training data, codes, and model weights are available in our repository \href{https://github.com/jeonchangbin49/De-limiter}{\texttt{{https://github.com/jeonchangbin49/De-limiter}}}.

\end{abstract}


\section{Introduction}
\label{sec:intro}
\vspace{-0.2cm}
The loudness war refers to an enduring phenomenon within the music industry, particularly in mixing and mastering, to excessively raise the overall loudness of music \cite{vickers2010loudness}. This trend is based on the common belief among artists and engineers that "louder is better" \cite{vickers2010loudness}. To increase the final loudness of music, producers and engineers use various signal processing tools, with a limiter being one of the most important tools for reducing dynamic range and increasing loudness \cite{stikvoort1986digital}.

However, many researchers and music enthusiasts argue that excessively raising the loudness of music using limiters actually harms the quality of music. In particular, Croghan, et al. \cite{croghan2012quality} and Campbell, et al. \cite{campbell2017listener} have shown that audio with excessive compression sounds worse than audio without any compression or moderate compression. Additionally, excessive compression and loudness can cause ear fatigue and hearing loss \cite{basner2014auditory, world2015hearing}. As a result, many music streaming sites now utilize loudness normalization to provide users with a standardized volume of music playback \cite{spotify_loudness, sage_loudness}.
 
Despite these research findings and trends, many artists and engineers continue to excessively raise the loudness of music with limiters \cite{dredge2013pop, milner2019they, haghbayan2020temporal}. The fatal drawback of applying a limiter to increase loudness is that it is a non-invertible transformation, which makes the de-limiter task difficult.

On the other hand, the recent remarkable advancements in deep neural networks have demonstrated outstanding performance in diverse tasks with irreversible and ill-defined characteristics. 
Generative models are representative examples, such as Text-to-Speech \cite{oord2016wavenet, wang2017tacotron}, Text-to-Audio \cite{kreuk2022audiogen}, Text-to-Music \cite{agostinelli2023musiclm} synthesis models, which generate audio outputs according to given sentences.
Given these advancements, we conjecture that neural networks can play a key role in building the de-limiter application.

In this paper, we introduce music de-limiter networks that can restore music with high loudness to its original state.
To this end, we propose the inverse dynamic range compression framework, namely \textit{sample-wise gain inversion (SGI)}, which estimates sample-wise gain values that invert the original signal from given compressed signal.
By this approach, we restore uncompressed music that has no audible artifacts and phase errors.
Furthermore, we present \textit{musdb-XL-train} dataset, which consists of 300k commercial limiter-applied music segments made for training de-limiter networks that are suitable for real-world applications.

Our research is designed not only to protect listeners' hearing but also to provide a richer and more diverse listening experience.
Especially, we anticipate that our research will have practical applications in commercial streaming services, enhancing the overall musical experience for listeners. 
Also, our de-limiter will be particularly beneficial to music producers and artists who use music sampling techniques, as it allows them to work with source material that closely resembles the original uncompressed signal, thus providing greater creative freedom.

\section{Related Works}
\label{sec:related_works}
\vspace{-0.2cm}
\textbf{De-compression} with known reverse-processing parameters has been proposed in \cite{lachaise2008inverting, gorlow2013model}. However, since this approach requires the parameters used in compression, it is unworkable to apply this on many of real-world music, where such parameters are unavailable. Instead, we propose a data-driven approach for inversion of a limiter.

\textbf{De-clipping}, which is a task of restoring clean audio from clipped signal, shares a similar attribute with de-compression because both clipping and compression are non-linear operations. It has been investigated using iterative methods \cite{zavivska2020survey, janssen1986adaptive, gaultier2021sparsity} or DNN-based methods \cite{kashani2019image, mack2019declipping, tanaka2022applade}. 
Recently, \cite{imort2022distortion} introduced distortion audio effect removal models using architectures that were proposed for source separation.
Also, \cite{moliner2023solving} proposed the generative approach that used a diffusion model for de-clipping. However, prior studies have not demonstrated that other audio effects like compression can be solved by DNN-based approach. 

In this paper, we perform evaluations on various frameworks for de-limiter; \textit{\romannumeral1)} synthesis (in Fig. \ref{fig:proposed} (a)), which showed great performance in recent music source separation \cite{defossez2019music, choi2019investigating}, \textit{\romannumeral2)} masking (in Fig. \ref{fig:proposed} (b)), which became a general framework in recent speech separation \cite{luo2019conv} and \textit{\romannumeral3)} our SGI method (in Fig. \ref{fig:proposed} (c)).

\section{Proposed Methods and Dataset}
\label{sec:methods}
\vspace{-0.3cm}
\subsection{Sample-wise Gain Inversion}
\vspace{-0.3cm}
A dynamic range compressor (or a compressor) works by reducing the gain of given signal when it exceeds the pre-defined \textit{threshold}, by the amount of \textit{ratio} parameter with given \textit{attack} and \textit{release} time. A limiter is a compressor that has an infinite ratio parameter \cite{zolzer2002dafx}. Though it is considered as a non-linear time-varying processor, it can be regarded as a simple sample-wise (or element-wise) linear gain manipulator \cite{zolzer2002dafx}; if we know the gain manipulation parameters for every single sample of given audio waveform, we can accurately obtain the dynamic range compressed output.

Based on this concept, we propose a \textit{sample-wise gain inversion} (SGI) framework for the de-limiter application. Given the dynamic range compressed signal, the de-limiter estimates its sample-wise gain inversion parameters. Each parameter has a range from 0 to 1. With such parameters, the original uncompressed signal can be simply obtained by element-wise multiplication in time-domain. This framework for the de-limiter has several following advantages. 

\textbf{Minimal artifacts.} Instead of directly generating waveforms using neural networks, since we manipulate sample-wise gain of given limiter-applied signal, it is much easier to avoid audible artifacts.

\textbf{Parallel mix.} Since SGI does not introduce any phase errors, we introduce \textit{parallel mix}, which is a technique that linearly mixes the given heavily compressed signal and the de-limiter network output properly. This is inspired by a music production technique called \textit{parallel compression}. Although heavy compression induces perceptual quality degradation, the decision of applying such compression is made by professional engineers, producers, and artists, which means it still has high sound quality. Furthermore, it can be perceptually worse to listen to purely uncompressed music in comparison to its moderately compressed version \cite{croghan2012quality}. Parallel mix can be a simple yet effective solution for such problems while giving maximum options to listeners.

\textbf{Light-weight network.} 
Even in heavily compressed songs with very high loudness, it is rare for the limiter to be engaged throughout the entire duration of the song from start to finish. A significant portion of the samples remain uncompressed. In other words, the proposed SGI de-limiter network estimates sample-wise values ranging from 0 to 1, with the majority of them expected to be close to a specific value for uncompressed sections.
This allows us to design shallow networks that can achieve excellent performance while maintaining fast training and inference speed.

\begin{figure}[t]
  \centering
  \centerline{\includegraphics[width=8.0cm]{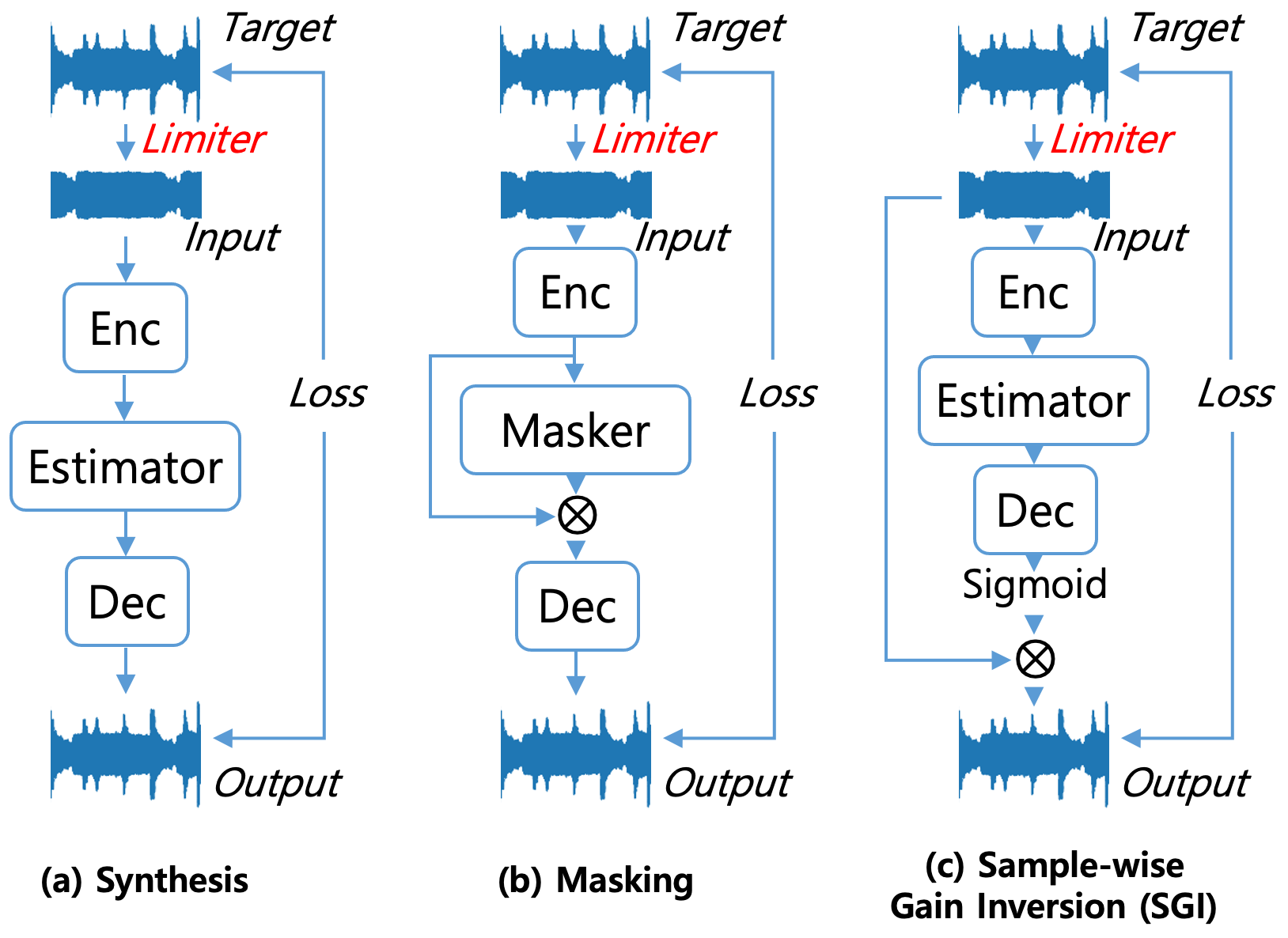}}
  \vspace{-0.2cm}
  \caption{Proposed de-limiter frameworks.}
  \label{fig:proposed}
\end{figure}

\vspace{-0.2cm}
\subsection{Architectures and Objectives}
\vspace{-0.2cm}
We borrowed the \textit{Conv-TasNet} \cite{luo2019conv} architecture, which shows good performance on audio source separation tasks, with simple modifications on the design.
The original Conv-TasNet follows the masking method on the learnable-basis, which is the element-wise multiplication between encoder and masker outputs in Fig. \ref{fig:proposed} (b).

However, for the SGI framework, we modify the activation of decoder outputs to sigmoid so that it can estimate gain inversion parameters in range of 0 to 1. Then, the estimated parameters are directly multiplied on the input signal. That is, the masking is applied on the input waveform and the decoder outputs as depicted in Fig. \ref{fig:proposed} (c), instead of the masking on encoder and masker outputs in Fig. \ref{fig:proposed} (b).

We use scale-invariant source-to-distortion ratio (SI-SDR) \cite{le2019sdr} as a training objective for the training stability. Note that the network output possibly has different scale with its input so we manually manipulate the scale of the output by loudness normalization in the inference stage following many streaming services' default configurations.

\vspace{-0.2cm}
\subsection{\textit{Musdb-XL-train} data}
\vspace{-0.2cm}
Training data is an essential part of modern deep learning. Since there exist no public data for training the de-limiter network (pairs of heavily compressed music and its original state), we introduce two ways of obtaining data.

\textbf{JUCE} \cite{juce} is an open-source C++ framework for an audio application. We use \textit{JUCE} for making training data of de-limiter networks on-the-fly with random gain scaling and mixing \cite{uhlich2017improving} of four stems of \textit{musdb-HQ} \cite{musdb18-hq}.

\textbf{Commercial digital plug-ins} are widely used in modern mastering process. However, it is generally infeasible to use them on-the-fly during training data construction because many popular plug-ins do not support Linux operating systems, which is widely used in training environments for neural networks such as Ubuntu. Furthermore, even if it is possible, not all researchers can easily use them because they are not free. One might suggest to imitate and implement such limiter plug-ins for an on-the-fly data sampler but their specific algorithms are confidential in general. To overcome these, we  introduce \textit{musdb-XL-train}, the training data for the de-limiter application built with the commercial limiter plug-in. Specifically, we made 4 seconds of 300,000 limiter-applied segments, where the original mixture is obtained by random mixing of \textit{musdb-HQ} train data \cite{uhlich2017improving}. We also made the limiter-applied version of 100 original training data without random mixing. The iZotope Ozone 9 Maximizer with random parameters was used for the data construction. The detailed explanations and data itself are publicly available in our code repository.

\begin{table*}[t]\centering
\resizebox{\columnwidth}{!}{
\begin{tabular}{llcccccccc}\toprule
Train data                                                                    & Methods              & \begin{tabular}[c]{@{}c@{}}Layer size\\ (X, R)\end{tabular} & \begin{tabular}[c]{@{}c@{}}SI-SDR $\uparrow$ \\ {[}dB{]}\end{tabular} & \begin{tabular}[c]{@{}c@{}}Multi-spec\\ MSE $\downarrow$\end{tabular} & \begin{tabular}[c]{@{}c@{}}PEAQ $\uparrow$\\ {[}ODG{]}\end{tabular} & FAD $\downarrow$  & \begin{tabular}[c]{@{}c@{}}\# of\\ params. $\downarrow$\end{tabular} & MACs $\downarrow$ & \begin{tabular}[c]{@{}c@{}}SI-SDR $\uparrow$ \\ w/ prl. mix\end{tabular}\\ \midrule
- & \textit{musdb-HQ} vs. \textit{XL} & - & 20.4 & 0.26 & 0.215 & 0.015 & - & - & -\\ \midrule
\multirow{6}{*}{\textit{JUCE}}                                                   & \textit{1-s)} Synthesis      & (5,2)                                                       & 4.7                                                        & 6.54                                                     & -3.124                                                    & 21.12 &     2.41M                                                    &  2,873G & -5.0    \\
                                                                              & \textit{1-d)} Synthesis      & (8,3)                                                       & 5.2                                                        & 6.03                                                     & -2.822                                                    & 0.727 &   5.23M                                                      & 2,987G & -13.7     \\ \cmidrule{2-10}
                                                                              & \textit{2-s)} Masking        & (5,2)                                                       & 9.4                                                        & 3.45                                                     & -1.026                                                    & 2.931 & 2.41M                                                        &  2,873G & -1.8   \\ 
                                                                              & \textit{2-d)} Masking        & (8,3)                                                       & 11.1                                                       & 2.54                                                     & -0.672                                                    & 0.198 &   5.23M                                                      & 2,987G & -19.1     \\ \cmidrule{2-10}
                                                                              & \textit{3-s)} SGI (Proposed) & (5,2)                                                       & 9.5                                                        & 3.36                                                     & -0.584                                                    & 0.384 & \textbf{2.35M}                                                        &   \textbf{1,483G} & 15.7  \\ 
                                                                              & \textit{3-d)} SGI (Proposed) & (8,3)                                                       &          9.8                                                  &     3.16                                                     &  -0.500                                                         & 0.351      & 5.17M                                                        &  1,597G & 15.9   \\ \midrule
\multirow{6}{*}{\begin{tabular}[c]{@{}l@{}}\textit{Musdb-XL}\textit{-train}\\(Proposed)\end{tabular}} & 
\textit{4-s)} Synthesis
& (5,2)                                                       & 12.1                                                       & 2.96                                                     & -1.703                                                    & 0.210 &   2.41M                                                      &   2,873G  & -9.2 \\
                                                                              &
                                                                              
                                                                              \textit{4-d)} Synthesis
                                                                              & (8,3)                                                       & 12.7                                                       & 2.74                                                     & -1.498                                                    & 0.186 &  5.23M                                                       & 2,987G & -13.5    \\ \cmidrule{2-10}
                                                                              &
                                                                              
                                                                              \textit{5-s)} Masking
                                                                              & (5,2)                                                       & 23.9                                                       & 0.11                                                     & -0.282                                                    & 0.004 & 2.41M                                                        & 2,873G  & 23.2   \\
                                                                              &
                                                                              
                                                                              \textit{5-d)} Masking
                                                                              & (8,3)                                                       & \textbf{24.5}                                                       & \textbf{0.10}                                                     & -0.320                                                    & 2.670 &  5.23M                                                       &  2,987G & 4.7   \\ \cmidrule{2-10}
                                                                              &
                                                                              \textit{6-s)} SGI (Proposed)
                                                                              & (5,2)                                                       & 24.0                                                       & 0.12                                                     & -0.154                                                    & 0.003 & \textbf{2.35M}                                                        & \textbf{1,483G} & 22.9     \\ 
                                                                              &
                                                                              \textit{6-d)} SGI (Proposed)
                                                                              & (8,3)                                                       & 24.4                                                           &   0.10                                                       & \textbf{-0.081}                                                          &  \textbf{0.002}     &  5.17M                                                       & 1,597G & \textbf{23.5}    \\\bottomrule
\end{tabular}
}
\vspace{-0.2cm}
\caption{
Evaluation results between \textit{musdb-HQ} and the de-limiter outputs given \textit{musdb-XL} data. The last column indicates the results when using the parallel mix technique. The first row exceptively indicates the comparison between the \textit{musdb-HQ} and \textit{musdb-XL} for a baseline. 
} \label{tab:performance_sgi}
\end{table*}

\vspace{-0.3cm}
\section{Experiments}
\label{sec:experiments}
\vspace{-0.3cm}
\subsection{Setup and Metrics}
\vspace{-0.3cm}
For the evaluations, we use the \textit{musdb-XL} \cite{jeon2022towards}, the mastering-finished version (in terms of a limiter) of the \textit{musdb-HQ} test subset, as input for de-limiter networks, and compare the de-limited output with corresponding \textit{musdb-HQ} data.
We compare \textit{shallower (s)} and \textit{deeper (d)} models for each framework. They have different number of repeat (R in Table \ref{tab:performance_sgi}) and each repeat has different convolutional block size (X in Table \ref{tab:performance_sgi}) of the Conv-TasNet \cite{luo2019conv}. The \textit{shallower} model has a 183ms of receptive field and the \textit{deeper} model has 2,223ms. 
The detailed training settings is contained in our codes.
% shallower => 125
% 125 * 64 + 64 = 8064 => 183ms
% deeper => 1531 
% 1531 * 64 + 64 = 98048 => 2.223s

All of the evaluations are performed in loudness normalized condition (-14 LUFS) in accordance with various modern streaming services \cite{spotify_loudness, sage_loudness}. Loudness calculation is performed with \texttt{pyloudnorm} \cite{steinmetz2021pyloudnorm}. Also, we apply a JUCE limiter using \texttt{pedalboard} \cite{sobot_peter_2023_7817838}, a python wrapper of JUCE.

As objective metrics, we measure SI-SDR, mean squared error (MSE) of multi-resolution magnitude spectrograms, perceptual evaluations of audio quality (PEAQ) scores \cite{thiede2000peaq}, Fr{\'e}chet audio distance (FAD) \cite{kilgour2019frechet}, the number of network parameters, and the multiply-accumulate operations (MACs) given 1-minute stereo audio input. We also report various dynamic parameters, such as the average of root mean square (RMS), crest factor, dynamic complexity \cite{streich2006music}, loudness range (LRA) \cite{ebu2011loudness}, and spectral centroid on each \textit{musdb-HQ} test subset, \textit{musdb-XL}, and the de-limiter outputs.

\vspace{-0.3cm}
\subsection{Objective Evaluations}
\label{sec:obj_eval}
\vspace{-0.3cm}

In Table \ref{tab:performance_sgi},
we confirm that our proposed SGI framework trained with the proposed \textit{musdb-XL-train} data (\textit{6-s)}) outperforms other shallower versions of synthesis- and masking-based de-limiters in every measure except multi-spec MSE.
Also, despite being only 45\% of the network size, it is remarkable that the \textit{shallower} \textit{6-s)} SGI model shows competitive performance with the \textit{deeper} \textit{6-d)} model in every score.
 This implies that an inductive bias of the SGI framework is powerful that it does not require huge size network to train a quality de-limiter. We consider this aspect will be especially useful for a future usage of the de-limiter in music streaming services or other applications which requires minimum possible resources for deployments.

Furthermore, it is noteworthy that the SI-SDR score of the masking-based \textit{5-d)} model is better than the SGI de-limiter \textit{6-d)}, while it fails badly on FAD score (2.670) whereas the \textit{6-d)} model has FAD score close to 0 (0.002). This is because of the phase error caused by the masking framework. The left and right channels of the outputs from the \textit{5-d)} model had the opposite polarity to the each other one. However, because the FAD calculation uses an input mono-channel audio of 16 kHz sample rate instead of full-bandwidth of 44.1 kHz sample rate, huge energy of waveforms are missing while converting stereo signal to mono (summation of the left and right channels and division by 2). Nevertheless, it is quite surprising that the same masking model with smaller network \textit{5-s)} actually performs better than the \textit{5-d)} in terms of FAD and shows no phase issues.

We also observe that the proposed \textit{musdb-XL-train} data works better than the training data made on-the-fly using \textit{JUCE}. Since a \textit{JUCE} limiter is implemented with two consecutive compressors without look-ahead feature, we assume that train data generated by \textit{JUCE} has different overall distribution compared to music made by applying commercial look-ahead limiter plug-ins. Qualitatively, we have also confirmed that such models suffer from audible volume fluctuation where kick drums exist, which makes outputs highly unnatural.

In Table \ref{tab:stats_sgi}, we notice that \textit{musdb-HQ} has higher crest factor than \textit{musdb-XL} while they have similar RMS values. This implies that although they have similar overall energy in loudness normalized condition, instantaneous peak is more prominent in an uncompressed version, \textit{musdb-HQ}. This characteristic is also related to the higher dynamic complexity and loudness range of \textit{musdb-HQ}. Furthermore, we notice that the spectral centroid of \textit{musdb-XL} is slightly higher than \textit{musdb-HQ}. Due to the non-linear nature of a limiter, which introduces harmonic distortions that enhance higher frequencies of given input signal, we assume this is a natural behavior.

\begin{table}
\resizebox{\columnwidth}{!}{
\begin{tabular}{lccccc} \toprule
\begin{tabular}[l]{@{}l@{}}Data // \\ Methods\end{tabular}                                                             & RMS   & \begin{tabular}[c]{@{}c@{}}Crest\\ factor\end{tabular} & \begin{tabular}[c]{@{}c@{}}Dynamic\\ complexity {[}dB{]}\end{tabular} & \begin{tabular}[c]{@{}c@{}}LRA\\ {[}LU{]}\end{tabular}  & \begin{tabular}[c]{@{}c@{}}Spectral\\ centroid {[}Hz{]}\end{tabular} \\ \midrule
\textit{musdb-HQ}                                                         & 0.150 & 6.73                                                   & 4.86                                                         & 7.48 & 1,464                                                       \\
\textit{musdb-XL}                                                         & 0.149 & 3.66                                                   & 4.66                                                         & 7.02 & 1,489                                                                                          \\ \midrule
\textit{6-s)} SGI
& 0.143 & 6.48                                                   & 4.84                                                         & 7.40 & 1,456       \\
\textit{6-d)} SGI
& 0.150 & 6.35                                                   & 4.86                                                         & 7.48 & 1,452                                                                                                          \\ \bottomrule   
\end{tabular}
}
\vspace{-0.2cm}
  \caption{Dynamic and spectral parameters of test data and the de-limiter outputs.}
  \label{tab:stats_sgi}
\end{table}

\vspace{-0.3cm}
\subsection{Stem-wise Analysis}
\label{sec:stem_analysis}
\vspace{-0.3cm}

In Table \ref{tab:stemwise}, the comparison was performed on each stem (\textit{vocals}, \textit{bass}, \textit{drums}, and \textit{other}) of \textit{musdb-HQ} and the de-limiter outputs. Exceptively, the first row shows the results on comparison between the \textit{musdb-HQ} and \textit{musdb-XL} to analyze how limiter-applied signal is different from its original state stem-wisely. To obtain the stem of de-limiter outputs in this section, we first calculate sample-wise gain ratio between specific \textit{musdb-XL} data and its de-limiter output, then multiply it to each stem of \textit{musdb-XL} data following the method used in \textit{LimitAug} \cite{jeon2022towards}.
We use SI-SDR and the median of the dynamic complexity difference between each song from \textit{musdb-HQ} and its de-limiter output for the stem-wise comparison. 

In general, \textit{drums} are main elements that trigger an operation of a limiter because they consist of sources like a kick or snare, which has high instantaneous energy. Hence, it is possible to assume that distortions caused by a limiter will be most prominent in \textit{drums} than other sources. In the first row of Table \ref{tab:stemwise}, we quantitatively confirm that \textit{drums} are the most distorted sources by a limiter and also confirm that their dynamic complexity is highly reduced by a limiter than any other stems. Surprisingly, as can be seen in
3.4
dB increases in SI-SDR, our de-limiter networks show great performance on restoring such distortions on \textit{drums}. Also, we observe that our models greatly reduces the $\Delta$ dynamic complexity of \textit{drums} from -0.20 to
0.02
dB.

\vspace{-0.3cm}
\subsection{Analysis on Parallel Mix}
\label{sec:parallel}
\vspace{-0.3cm}
We conduct additional analysis by calculating SI-SDR between \textit{musdb-HQ} and a \textit{parallel mix} output, which is a linear mixture of specific \textit{musdb-XL} data and its de-limiter output. Note that we used a parallel mix of each audio with equal ratio (0.5:0.5) but the ratio can be freely chosen for further use cases.
As shown in the last column of Table \ref{tab:performance_sgi}, when we make a parallel mix using the shallower masking-based de-limiter, the \textit{5-s)} model,  SI-SDR is 23.2 dB, whereas the deeper model \textit{5-d)} shows 4.7 dB. This implies that the masking-based de-limiter does not guarantee the phase-aligned outputs, unlike the proposed SGI de-limiter, where SI-SDR scores are 22.9 (\textit{6-s)}) and 23.5 (\textit{6-d)}) dB.
With this characteristic, it is possible to build an application that users can adjust the linear mix coefficient of original music and de-limited music, which will give listeners maximum options for better listening experiences.

\begin{table}
\resizebox{8.5cm}{!}{
\begin{tabular}{@{\hspace{1pt}}l@{\hspace{6.5pt}}c@{\hspace{6.5pt}}c@{\hspace{6.5pt}}c@{\hspace{6.5pt}}cc@{\hspace{6.5pt}}c@{\hspace{6.5pt}}c@{\hspace{6.5pt}}c@{\hspace{2pt}}}
\toprule
\multirow{2}{*}{\begin{tabular}[c]{@{}l@{}}Data // \\ Methods\end{tabular}}        & \multicolumn{4}{c}{SI-SDR $\uparrow$ {[}dB{]}}                               & \multicolumn{4}{c}{$\Delta$ Dyn. complexity $|\downarrow|$ {[}dB{]}}           \\ \cmidrule(r){2-5} \cmidrule(r){6-9}
                                                                 & \textit{vocals} & \textit{bass} & \textit{drums} & \textit{other} & \textit{vocals} & \textit{bass} & \textit{drums} & \textit{other} \\ \midrule
\textit{musdb-XL} & 23.7 & 25.1 & 20.0 & 25.5 & -0.07 & -0.02 & -0.20 & -0.05\\ \midrule
                                    \textit{6-s)} SGI
                                    &  26.7 & 28.1 & 23.4 & 28.5 & \textbf{0.04} & \textbf{0.03} & \textbf{0.02} & \textbf{0.02}               \\
                                    \textit{6-d)} SGI
                                    & \textbf{26.9}                & \textbf{28.4}              & \textbf{23.5}               &             \textbf{28.7}   & 0.12                &  0.05             &              0.07  & 0.03               \\ \bottomrule
\end{tabular}
}
\vspace{-0.2cm}
  \caption{Stem-wise analysis of the proposed de-limiter networks. For $\Delta$ dynamic complexity, the lower absolute value, the better.}
  \label{tab:stemwise}
\end{table}

\begin{table}
\resizebox{3.6cm}{!}{
\begin{tabular}{@{\hspace{1pt}}l@{\hspace{8pt}}c@{\hspace{8pt}}c@{\hspace{1pt}}}
\toprule
Methods & \begin{tabular}[c]{@{}c@{}}SI-SDR\\ {[}dB{]}\end{tabular}  & \begin{tabular}[c]{@{}c@{}}LRA\\ {[}LU{]}\end{tabular}  \\ \midrule
GLN
& 24.0            & 7.40         \\ \midrule
LN
& 23.3            & 7.24         \\
BN
& 24.0            & 7.29         \\
FGLN
& \textbf{24.0}            & \textbf{7.41}         \\ \bottomrule
\end{tabular}
}
\vspace{-0.2cm}
  \caption{Results when using different normalization layers.}
  \label{tab:noramlization}
\end{table}

\vspace{-0.3cm}
\subsection{Analysis on Real-World Music and Architecture Design}
\label{sec:robustness}
\vspace{-0.3cm}
Since both \textit{musdb-XL-train} and \textit{musdb-XL} (test) are made with the same commercial limiter, one might raise a question about out-of-domain performance of the proposed models. 
Hence, to check the robustness of the proposed de-limiter networks, we qualitatively conduct the analysis using various real-world popular music. 
As a representative example, we apply our de-limiter on \textit{Metallica}'s \textit{My Apocalypse}, which is the song from the album \textit{Death Magnetic (released in 2008)}. This album had once gained lots of criticisms for heavy compression \cite{michaels2008metallica}. As can be seen in Fig. \ref{fig:spec}, our de-limiter has successfully estimated the plausible de-limited signal of the song. The de-limited signal shows much frequent peaks in waveform and more audible presence in kick and snare drums as shown in the spectrograms. We have observed this behavior in various genres of music and they are available on our demo page.
 
We have also discovered that our models are especially proficient at increasing the overall dynamic range, the loudness difference between relatively quieter parts (e.g., verse) and louder parts (e.g., chorus) of given long music signal. 
This is remarkable because our models have never trained with few-minutes long signal but only trained with 4-second segments. We suspect this characteristic comes from the global layer normalization (GLN) \cite{luo2019conv} used in the \textit{Conv-TasNet}, which normalizes values of whole input representations. We perform a few additional experiments with replacing GLN of \textit{6-s)} model to batch normalization (BN) \cite{ioffe2015batch}, layer normalization (LN) \cite{ba2016layer}, and lastly, feature-wise global layer normalization (FGLN) \cite{kavalerov2019universal} that also normalizes values of whole input representations like GLN but feature-wisely.

In Table \ref{tab:noramlization}, we discover that all of the normalization layers show high performance but the loudness ranges of LN and BN were slightly lower than GLN and FGLN. This implies that the characteristic of GLN and FGLN is especially useful for increasing the overall dynamic range of music while reflect the context of whole song. We insist that this feature should be considered for future work for designing better architecture on de-limiter networks.

\begin{figure}[t]
  \centering
  \vspace{-0.3cm}
  \centerline{\includegraphics[width=\columnwidth]{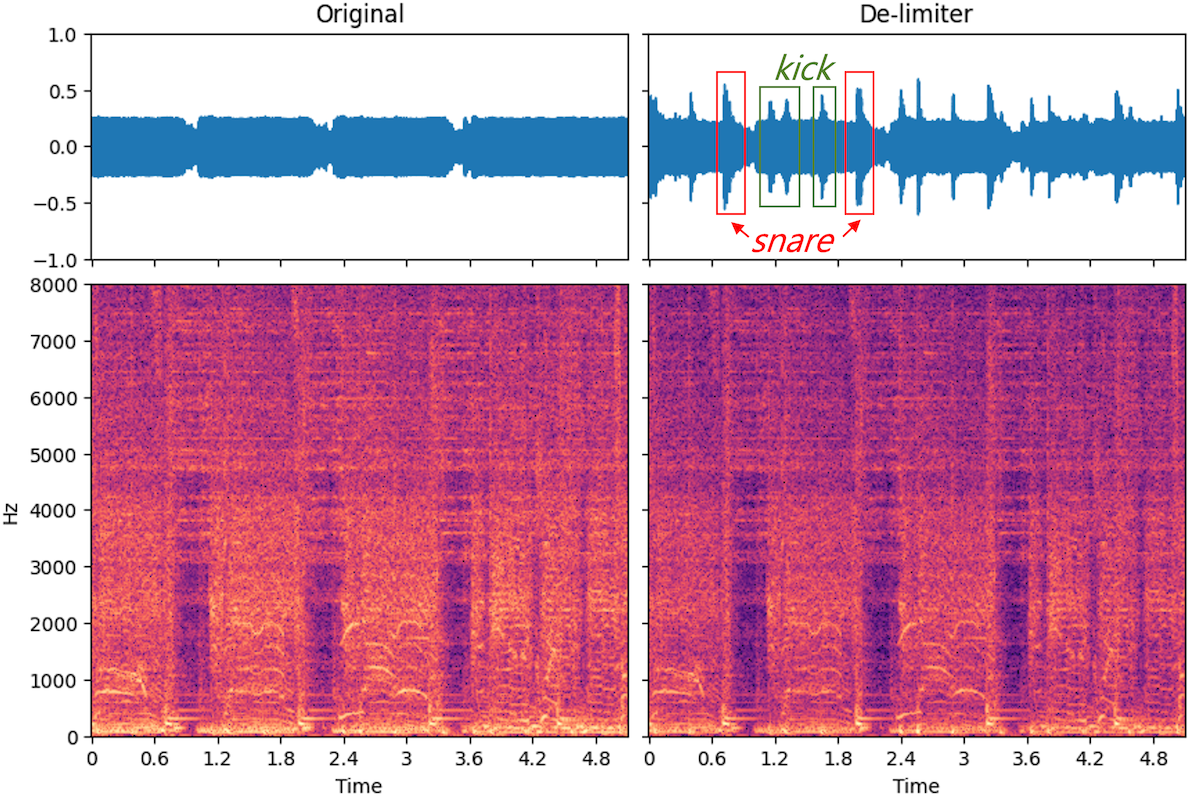}}
  \vspace{-0.3cm}
  \caption{The de-limiter example on \textit{Metallica - My Apocalypse}.}
  \label{fig:spec}
\end{figure}

\vspace{-0.3cm}
\section{Conclusions}
\label{sec:conclusions}
\vspace{-0.3cm}
We have proposed music de-limiter networks based on the \textit{sample-wise gain inversion} (SGI) framework to enhance the listening and creation experience of music. Additionally, we have introduced the \textit{musdb-XL-train} dataset to train de-limiter networks. Our proposed de-limiter networks trained with the \textit{musdb-XL-train} dataset have exhibited outstanding performance in de-limiting \textit{musdb-XL} to its original state, \textit{musdb-HQ}.
Our exploration of the SGI framework has revealed several advantages, including the absence of audible artifacts unlike recent generation-based neural networks, no phase errors, which enables the use of \textit{parallel mix}, and a light-weight network design that exhibits competitive performance with twice as deeper networks. We strongly believe that our proposed de-limiter frameworks and dataset can offer novel experiences in music listening and creation.

\vspace{-0.5cm}
\section{Acknowledgements}
\vspace{-0.3cm}
We are grateful to Marco A. Mart{\'\i}nez-Ram{\'\i}rez for his fruitful comments on the evaluations of our proposed models.

\vspace{-0.4cm}
\bibliographystyle{IEEEtran}
\bibliography{refs23}

\begin{thebibliography}{10}
\providecommand{\url}[1]{#1}
\def\UrlFont{\rmfamily}
\providecommand{\newblock}{\relax}
\providecommand{\bibinfo}[2]{#2}
\providecommand\BIBentrySTDinterwordspacing{\spaceskip=0pt\relax}
\providecommand\BIBentryALTinterwordstretchfactor{4}
\providecommand\BIBentryALTinterwordspacing{\spaceskip=\fontdimen2\font plus
\BIBentryALTinterwordstretchfactor\fontdimen3\font minus \fontdimen4\font\relax}
\providecommand\BIBforeignlanguage[2]{{%
\expandafter\ifx\csname l@#1\endcsname\relax
\typeout{** WARNING: IEEEtran.bst: No hyphenation pattern has been}%
\typeout{** loaded for the language `#1'. Using the pattern for}%
\typeout{** the default language instead.}%
\else
\language=\csname l@#1\endcsname
\fi
#2}}

\bibitem{vickers2010loudness}
E.~Vickers, ``The loudness war: Background, speculation, and recommendations,'' in \emph{AES Convention 129}.\hskip 1em plus 0.5em minus 0.4em\relax AES, 2010.

\bibitem{stikvoort1986digital}
E.~F. Stikvoort, ``Digital dynamic range compressor for audio,'' \emph{Journal of the AES (JAES)}, vol.~34, no. 1/2, pp. 3--9, 1986.

\bibitem{croghan2012quality}
N.~B. Croghan, K.~H. Arehart, and J.~M. Kates, ``Quality and loudness judgments for music subjected to compression limiting,'' \emph{The Journal of the Acoustical Society of America (ASA)}, vol. 132, no.~2, pp. 1177--1188, 2012.

\bibitem{campbell2017listener}
W.~Campbell, J.~Paterson, and I.~Van~der Linde, ``Listener preferences for alternative dynamic-range-compressed audio configurations,'' \emph{JAES}, vol.~65, no. 7/8, pp. 540--551, 2017.

\bibitem{basner2014auditory}
M.~Basner, W.~Babisch, A.~Davis, M.~Brink, C.~Clark, S.~Janssen, and S.~Stansfeld, ``Auditory and non-auditory effects of noise on health,'' \emph{The lancet}, vol. 383, no. 9925, pp. 1325--1332, 2014.

\bibitem{world2015hearing}
{World Health Organization}, ``Hearing loss due to recreational exposure to loud sounds: a review,'' 2015.

\bibitem{spotify_loudness}
Spotify, ``Loudness normalization,'' Retrieved 14 Apr. 2023.

\bibitem{sage_loudness}
{Sage Audio}, ``Mastering for streaming: Platform loudness and normalization explained,'' Retrieved 14 Apr. 2023.

\bibitem{dredge2013pop}
S.~Dredge, ``Pop music is louder, less acoustic and more energetic than in the 1950s,'' \emph{The Guardian}, 25 Nov. 2013.

\bibitem{milner2019they}
G.~Milner, ``They really don’t make music like they used to,'' \emph{The New York Times}, 7 Feb. 2019. Retrieved 14 Apr. 2023.

\bibitem{haghbayan2020temporal}
H.~Haghbayan, E.~A. Coomes, and D.~Curran, ``Temporal trends in the loudness of popular music over six decades,'' \emph{Journal of General Internal Medicine}, vol.~35, no.~1, pp. 394--395, 2020.

\bibitem{oord2016wavenet}
A.~v.~d. Oord, S.~Dieleman, H.~Zen, K.~Simonyan, O.~Vinyals, A.~Graves, N.~Kalchbrenner, A.~Senior, and K.~Kavukcuoglu, ``Wavenet: A generative model for raw audio,'' \emph{arXiv preprint arXiv:1609.03499}, 2016.

\bibitem{wang2017tacotron}
Y.~Wang, R.~Skerry-Ryan, D.~Stanton, Y.~Wu, R.~J. Weiss, N.~Jaitly, Z.~Yang, Y.~Xiao, Z.~Chen, S.~Bengio, \emph{et~al.}, ``Tacotron: Towards end-to-end speech synthesis,'' \emph{Proc. Interspeech}, pp. 4006--4010, 2017.

\bibitem{kreuk2022audiogen}
F.~Kreuk, G.~Synnaeve, A.~Polyak, U.~Singer, A.~D{\'e}fossez, J.~Copet, D.~Parikh, Y.~Taigman, and Y.~Adi, ``Audiogen: Textually guided audio generation,'' in \emph{The Eleventh International Conference on Learning Representations}, 2022.

\bibitem{agostinelli2023musiclm}
A.~Agostinelli, T.~I. Denk, Z.~Borsos, J.~Engel, M.~Verzetti, A.~Caillon, Q.~Huang, A.~Jansen, A.~Roberts, M.~Tagliasacchi, \emph{et~al.}, ``Musiclm: Generating music from text,'' \emph{arXiv preprint arXiv:2301.11325}, 2023.

\bibitem{lachaise2008inverting}
B.~Lachaise and L.~Daudet, ``Inverting dynamics compression with minimal side information,'' in \emph{International Conference on Digital Audio Effects (DAFx)}, 2008.

\bibitem{gorlow2013model}
S.~Gorlow and J.~D. Reiss, ``Model-based inversion of dynamic range compression,'' \emph{IEEE TASLP}, vol.~21, no.~7, pp. 1434--1444, 2013.

\bibitem{zavivska2020survey}
P.~Z{\'a}vi{\v{s}}ka, P.~Rajmic, A.~Ozerov, and L.~Rencker, ``A survey and an extensive evaluation of popular audio declipping methods,'' \emph{IEEE Journal of Selected Topics in Signal Processing}, vol.~15, no.~1, pp. 5--24, 2020.

\bibitem{janssen1986adaptive}
A.~Janssen, R.~Veldhuis, and L.~Vries, ``Adaptive interpolation of discrete-time signals that can be modeled as autoregressive processes,'' \emph{IEEE TASLP}, vol.~34, no.~2, pp. 317--330, 1986.

\bibitem{gaultier2021sparsity}
C.~Gaultier, S.~Kiti{\'c}, R.~Gribonval, and N.~Bertin, ``Sparsity-based audio declipping methods: Selected overview, new algorithms, and large-scale evaluation,'' \emph{IEEE TASLP}, vol.~29, pp. 1174--1187, 2021.

\bibitem{kashani2019image}
H.~B. Kashani, A.~Jodeiri, M.~M. Goodarzi, and S.~G. Firooz, ``Image to image translation based on convolutional neural network approach for speech declipping,'' \emph{arXiv preprint arXiv:1910.12116}, 2019.

\bibitem{mack2019declipping}
W.~Mack and E.~A. Habets, ``Declipping speech using deep filtering,'' in \emph{Proc. of WASPAA}.\hskip 1em plus 0.5em minus 0.4em\relax IEEE, 2019, pp. 200--204.

\bibitem{tanaka2022applade}
T.~Tanaka, K.~Yatabe, M.~Yasuda, and Y.~Oikawa, ``Applade: Adjustable plug-and-play audio declipper combining dnn with sparse optimization,'' in \emph{Proc. of ICASSP}.\hskip 1em plus 0.5em minus 0.4em\relax IEEE, 2022, pp. 1011--1015.

\bibitem{imort2022distortion}
J.~Imort, G.~Fabbro, M.~A. Mart{\'\i}nez-Ram{\'\i}rez, S.~Uhlich, Y.~Koyama, and Y.~Mitsufuji, ``Distortion audio effects: Learning how to recover the clean signal,'' in \emph{Proc. of ISMIR}, 2022.

\bibitem{moliner2023solving}
E.~Moliner, J.~Lehtinen, and V.~V{\"a}lim{\"a}ki, ``Solving audio inverse problems with a diffusion model,'' in \emph{ICASSP 2023-2023 IEEE International Conference on Acoustics, Speech and Signal Processing (ICASSP)}.\hskip 1em plus 0.5em minus 0.4em\relax IEEE, 2023, pp. 1--5.

\bibitem{defossez2019music}
A.~D{\'e}fossez, N.~Usunier, L.~Bottou, and F.~Bach, ``Music source separation in the waveform domain,'' \emph{arXiv preprint arXiv:1911.13254}, 2019.

\bibitem{choi2019investigating}
W.~Choi, M.~Kim, J.~Chung, D.~Lee, and S.~Jung, ``Investigating u-nets with various intermediate blocks for spectrogram-based singing voice separation,'' \emph{Proc. of ISMIR}, 2020.

\bibitem{luo2019conv}
Y.~Luo and N.~Mesgarani, ``Conv-tasnet: Surpassing ideal time--frequency magnitude masking for speech separation,'' \emph{IEEE/ACM TASLP}, vol.~27, no.~8, pp. 1256--1266, 2019.

\bibitem{zolzer2002dafx}
U.~Z{\"o}lzer, X.~Amatriain, D.~Arfib, J.~Bonada, G.~De~Poli, P.~Dutilleux, G.~Evangelista, F.~Keiler, A.~Loscos, D.~Rocchesso, \emph{et~al.}, \emph{DAFX-Digital audio effects}.\hskip 1em plus 0.5em minus 0.4em\relax John Wiley \& Sons, 2002.

\bibitem{le2019sdr}
J.~Le~Roux, S.~Wisdom, H.~Erdogan, and J.~R. Hershey, ``Sdr--half-baked or well done?'' in \emph{Proc. of ICASSP}.\hskip 1em plus 0.5em minus 0.4em\relax IEEE, 2019, pp. 626--630.

\bibitem{juce}
\BIBentryALTinterwordspacing
JUCE, ``Juce,'' 2022. [Online]. Available: \url{https://github.com/juce-framework/JUCE}
\BIBentrySTDinterwordspacing

\bibitem{uhlich2017improving}
S.~Uhlich, M.~Porcu, F.~Giron, M.~Enenkl, T.~Kemp, N.~Takahashi, and Y.~Mitsufuji, ``Improving music source separation based on deep neural networks through data augmentation and network blending,'' in \emph{Proc. of ICASSP}.\hskip 1em plus 0.5em minus 0.4em\relax IEEE, 2017, pp. 261--265.

\bibitem{musdb18-hq}
\BIBentryALTinterwordspacing
Z.~Rafii, A.~Liutkus, F.-R. Stöter, S.~I. Mimilakis, and R.~Bittner, ``Musdb18-hq - an uncompressed version of musdb18,'' Aug. 2019. [Online]. Available: \url{https://doi.org/10.5281/zenodo.3338373}
\BIBentrySTDinterwordspacing

\bibitem{jeon2022towards}
C.-B. Jeon and K.~Lee, ``Towards robust music source separation on loud commercial music,'' in \emph{Proc. of ISMIR}, 2022.

\bibitem{steinmetz2021pyloudnorm}
C.~J. Steinmetz and J.~Reiss, ``pyloudnorm: A simple yet flexible loudness meter in python,'' in \emph{AES Convention 150}, 2021.

\bibitem{sobot_peter_2023_7817838}
\BIBentryALTinterwordspacing
P.~Sobot, ``Pedalboard,'' July 2021. [Online]. Available: \url{https://doi.org/10.5281/zenodo.7817838}
\BIBentrySTDinterwordspacing

\bibitem{thiede2000peaq}
T.~Thiede, W.~C. Treurniet, R.~Bitto, C.~Schmidmer, T.~Sporer, J.~G. Beerends, and C.~Colomes, ``Peaq-the itu standard for objective measurement of perceived audio quality,'' \emph{JAES}, vol.~48, no. 1/2, pp. 3--29, 2000.

\bibitem{kilgour2019frechet}
K.~Kilgour, M.~Zuluaga, D.~Roblek, and M.~Sharifi, ``Fr{\'e}chet audio distance: A reference-free metric for evaluating music enhancement algorithms.'' in \emph{Proc. of INTERSPEECH}, 2019, pp. 2350--2354.

\bibitem{streich2006music}
S.~Streich \emph{et~al.}, \emph{Music complexity: a multi-faceted description of audio content}.\hskip 1em plus 0.5em minus 0.4em\relax Universitat Pompeu Fabra Barcelona, Spain, 2006.

\bibitem{ebu2011loudness}
R.~EBU-Recommendation, ``Loudness normalisation and permitted maximum level of audio signals,'' \emph{European Broadcasting Union}, 2011.

\bibitem{michaels2008metallica}
S.~Michaels, ``Metallica album latest victim in 'loudness war'?'' \emph{The Guardian}, 17 Sep. 2008. Retrieved 25 Apr. 2023.

\bibitem{ioffe2015batch}
S.~Ioffe and C.~Szegedy, ``Batch normalization: Accelerating deep network training by reducing internal covariate shift,'' in \emph{Proc. of ICML}.\hskip 1em plus 0.5em minus 0.4em\relax pmlr, 2015, pp. 448--456.

\bibitem{ba2016layer}
J.~L. Ba, J.~R. Kiros, and G.~E. Hinton, ``Layer normalization,'' \emph{arXiv preprint arXiv:1607.06450}, 2016.

\bibitem{kavalerov2019universal}
I.~Kavalerov, S.~Wisdom, H.~Erdogan, B.~Patton, K.~Wilson, J.~Le~Roux, and J.~R. Hershey, ``Universal sound separation,'' in \emph{Proc. of WASPAA}.\hskip 1em plus 0.5em minus 0.4em\relax IEEE, 2019, pp. 175--179.

\end{thebibliography}

\end{sloppy}
\end{document}